\documentclass[fleqn,10pt]{wlscirep}
\title{Genesis of charge orders in high temperature superconductors}

\author[1,2,*]{Wei-Lin Tu}
\author[2]{Ting-Kuo Lee}
\affil[1]{Department of Physics, National Taiwan University, Daan Taipei 10617, Taiwan}
\affil[2]{Institute of Physics, Academia Sinica, Nankang Taipei 11529, Taiwan}

\affil[*]{stevedue19@gmail.com}

\begin{abstract}
One of the most puzzling facts about cuprate high-temperature superconductors in the lightly doped regime is the coexistence of uniform superconductivity and/or antiferromagnetism with many low-energy charge-ordered states in a unidirectional charge density wave or a bidirectional checkerboard structure.  Recent experiments have discovered that these charge density waves exhibit different symmetries in their intra-unit-cell form factors for different cuprate families. Using a renormalized mean-field theory for a well-known, strongly correlated model of cuprates, we obtain a number of charge-ordered states with nearly degenerate energies without invoking special features of the Fermi surface. All of these self-consistent solutions have a pair density wave intertwined with a charge density wave and sometimes a spin density wave. Most of these states vanish in the underdoped regime, except for one with a large d-form factor that vanishes at approximately 19$\%$ doping of the holes, as reported by experiments. Furthermore, these states could be modified to have a global superconducting order, with  a nodal-like density of states at low energy. 
\end{abstract}
\begin{document}

\flushbottom
\maketitle
%
%
\thispagestyle{empty}

\section*{Introduction}

Ever since the discovery of the high-Tc superconductivity, many low-energy charge-ordered states in the cuprate have been discovered.  Neutron scattering experiments[1] first emphasised the doping dependence of incommensurate magnetic peaks associated with unidirectional magnetic patterns or stripes. Later, soft X-ray scattering[2] also confirmed the presence of charge orders with these stripes. However, these experiments were performed on the 214($La_{2-x}Sr_{x}CuO_{4}$) cuprate family. For other cuprate families,  the evidence for bond-centred unidirectional domains was found via scanning tunneling spectroscopy[3,4].  The charge density wave(CDW) order was also found to be induced by the external magnetic field[5]. Recently, more results regrading charge-ordered states [6-10], and electron-doped cuprates[11] have been reported. The periods of these CDW and their doping dependence are quite different for different cuprate families[10].  In addition to the unidirectional stripe pattern, some experiments have also reported the possible existence of a bidirectional charge-ordered checkerboard pattern[12,13]. The unidirectional charge-ordered states or stripes were found to have a dominant d-like symmetry for the intra-unit-cell form factor, measured on the two oxygen sites by using  the resonant elastic x-ray scattering method[14,15] and via scanning tunneling spectroscopy (STS)[16]. However, different families seem to prefer different symmetries[14,15].  In the STS experiments[17], the density waves disappeared above 19$\%$ hole doping. Furthermore, the observation of these CDW states having nodal-like local density of states(LDOS) at low energy but strong spatial variation at high energy in STS[3] strongly implies a new unconventional superconducting  state.  

\vspace{2mm}

The existence of these great varieties of charge-ordered states has created a great debate regarding whether the strong coupling Hubbard model or the $t-J$ model[18] is the proper basic Hamiltonian to describe the cuprates.  Many believe that these states ``compete" with the superconductivity[19] and that their origin  may reveal the fundamental understanding of the mechanism of high superconducting temperatures in cuprates. The recent detection of the d-form factor at an oxygen site instead of at a Cu site[14-16] also raises the question about the suitability of the effective one-band Hubbard or $t-J$ model and the validity of replacing the oxygen hole with a Zhang-Rice singlet[20], which effectively supports a simpler one-band model with Cu only. Allais \textsl{et al.}[23] proposed that the d-symmetry of these form factors, referred to as bond orders[21,22] because they are measured between the nearest neighbour Cu bonds, arise from the strong correlation but without other intertwined orders. Furthermore, there are also doubts regarding whether a strong correlation is present or even needed to understand of the superconducting mechanism[24]. However, the complexities of the phase diagram and some recent theoretical works have indicated the possibility of a new phase of matter, i.e., the pair density wave (PDW)[25-28], as discussed in detail in a recent review article[25]. The new states are considered to have intertwined orders of PDW and CDW or spin density waves(SDW)[25]. 

\vspace{2mm}

For quite some time, various calculations[29-39] on the Hubbard and $t-J$ type models have revealed low-energy intertwined states appearing as stripes or bidirectional charge-ordered states, such as checkerboard(CB).  However, these works usually involved different approximations and parameters, which often resulted in different types of charge-ordered patterns, and these studies were mostly concentrated at a hole concentration of $1/8$, which is the most notable concentration in early experiments. Hence, it is not clear if these results were the consequence of the invoked assumption or the approximation used, or if they are a generic results in the phase diagrams of cuprates. There were attempts to produce these CDWs or PDWs using a different approach, such as using a mean field theory to study a $t-J$-like model but taking the strong correlation as only a renormalization effect of dispersion[21,22,40,41]. A spin-fluctuation mediated mechanism to produce these states was also proposed for the spin-fermion model[42].  Recently, a novel mechanism of PDW was proposed, i.e., Amperean pairing[28], by using the gauge theory formulation of the resonating-valence-bond picture. In most of these approaches, the wave vectors or periods of the density waves are related to special features of the Fermi surface, including nesting, hot spots or regions with large density of states. However, the opposite doping dependence of CDW periods, observed for 214 and 123($YBa_2Cu_3O_{6+\delta}$) compounds[10], makes the Fermi surface scenario worrisome. 

\vspace{2mm}

Amid all this confusion, recent numerical progress achieved by using the infinite projected entangled-pair states(iPEPS) method[43], has provided us with a new clue. It was found that the $t-J$ model has several stripe states, with nearly degenerate energy as the uniform state and, with coexistent superconductivity and antiferromagnetism.  When the number of variational parameters is extrapolated to infinity, the authors concluded that the anti-phase stripe, which has no net pairing, has slightly higher energy than the in-phase stripe with a net pairing, which in turn, also has slightly higher energy than the uniform state. This result is very consistent with the result of variational Monte Carlo calculations[29] based on the concept of the resonating-valence-bond picture[18].  Furthermore, the result is also consistent with that of renormalized mean-field theory by using a generalised Gutzwiller approximation(GWA)[44] to treat the projection operator in the $t-J$ model[30,45]. Hence, the result provides strong support to more carefully examine the low energy states of the $t-J$ model with the variational approach using GWA. 

\vspace{2mm}

Here, we report our findings from a much more extensive examination of the renormalized mean-field theory prediction using the GWA for the hole-doped $t-J$ model. We find many unidirectional and bidirectional charge-ordered states with nearly degenerate energies as the uniform state, especially in the lightly doped regime; thus, it is a much more general phenomenon than previously thought. All of these states have intertwined orders of PDW, CDW and/or SDW. One of the CDW states, denoted as AP-CDW, reveals a bond order pattern with a much larger d-form factor than $s'$ symmetry, as found in the experiment[16] with BSCCO ($Bi_2Sr_2CaCu_2O_{8+x}$) and NaCCOC ($Ca_{2-x}Na_xCuO_2Cl_2$). Furthermore, just as in the experiment[17], it vanishes beyond 19$\%$ hole doping. However, not all these charge-ordered states have a dominant $d$-form factor. For example, a different CDW intertwined with SDW and PDW, which is the familiar stripe reported long ago for 214 [1,2,29,31,36], has a larger $s'$ form factor, as reported in the experiment[15]. We further show that this AP-CDW state could be easily altered to become a superconducting state with a global d-wave pairing symmetry, while locally, each bond does not have the perfect d-wave symmetry. Its spectra shows  a large spatial variation at higher energies but with a d-wave nodal-like LDOS near zero energy as seen in the experiments[2,17].

\section*{Results and Discussions}

As mentioned above, the variational approach has been quite effective at capturing the physics of the strong correlation present in the $t-J$ model. By using GWA, we can replace the strong constraint of forbidding the double occupancy of two holes on the same site in the variational wave function using Gutzwiller factors[32,33,44,45]. Then, one can use just mean field theory to find the various low energy states. Details about the calculation are discussed in the Methods section. 

\vspace{2mm}

In our mean field theory, there are four variational order  parameters. Besides the hole density $\delta_i$,  the  local spin moment $m_i^v$ provides the antiferromagnetic correlation, the  pair field $\Delta_{ij\sigma}^v$ represents  the local electron pairing order, and bond order $\chi_{ij\sigma}^v$ is just the kinetic hopping term, where $i$ is a site position and $ij$ is the nearest neighbour bond.  An iterative method is used to self-consistently solve the mean-field Hamiltonian $H_{MF}$(Eq. (S7) in the Supplementary Material(SM)) for all the parameters, of which there could be more than 60. The convergence is achieved for every order parameter if its value changes by less than $10^{-3}$ between successive iterations.  All the calculations are performed on a 16 by 16 square lattice. To obtain various charge orders, specific patterns of $\delta_i$, $m_i^v$, and $\Delta_{ij\sigma}^v$ are input as initial values. The bond orders $\chi_{ij\sigma}^v$ are always initially assumed to be uniform. In most cases, we will obtain only uniform solutions such as the d-wave superconducting(dSC) state and/or coexistent antiferromagnetic(dSC-AFM) state, but sometimes the states with charge-ordered patterns are found as a self-consistent solution. 

\subsection*{Charge-ordered Patterns}

In addition to the two uniform solutions of a dSC state and a dSC-AFM state, there are many non-uniform charge-ordered states. For simplicity, we shall first present those charge-ordered states with a period of four lattice spaces ($4a_0$), as listed in Table \ref{tab:example}. Both the pair field $\Delta_{ij\sigma}^v$ and the spin moment $m_i^v$ could have positive and negative values. It turns out that if there is a SDW or a bidirectional spin CB (sCB) present, then it always has a period of $8a_0$, with two domains of size $4a_0$ with opposite antiferromagnetic directions joining  together. The pair field has more choices. It could always be positive, with all of its x-bond pair field being positive and y-bond pair field being negative: thus, it would have a net total non-zero pair field. This is called an in-phase (IP) state, with a period of $4a_0$. However, just like the spin moment, the pair field could also have two domains with opposite signs and a domain wall in between: this state is known as the anti-phase (AP) state, with a period of $8a_0$.  Thus, we could have four possible states for each unidirectional CDW or bidirectional charge CB (cCB), as we either have an IP or AP pair field with or without SDW.  However, we only have three such states in Table \ref{tab:example} because we cannot find a solution with an IP pair field and CDW only. This result is due to the choice of the commensurate period being $4a_0$. Later, we will show a state with a net pairing order or IP pairing state and CDW, which occurs if we do not require solutions to be commensurate with the lattice. 

\vspace{2mm}

\begin{figure}[ht]
\centering
\includegraphics[width=\linewidth]{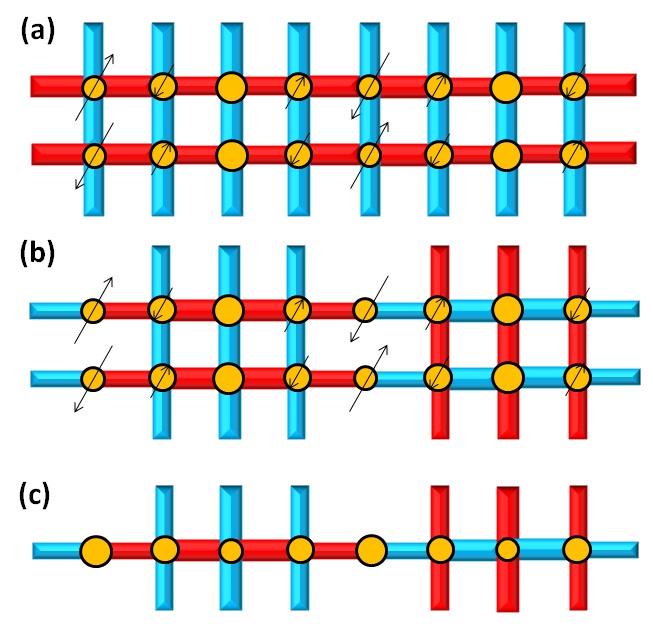}
\caption{}
\label{Fig12}
\end{figure}

Figure \ref{Fig12} shows a schematic illustration of the modulations of the pair field, charge density and spin moment  for the three stripes with hole concentration of 0.1. The magnitude of the pair field is proportional to the width of the bond; red(cyan) denotes positive(negative) value. The size of the arrow is proportional to the spin moment, and the size of the circle represents the hole density.  A similar figure for the three bidirectional CB patterns is shown in  Figure 7 in SM.  There is one domain wall corresponding to the vanishing spin moment for IP-CDW-SDW in Figure \ref{Fig12}a or the vanishing pair field for AP-CDW in Figure \ref{Fig12}c. Both domain walls are present for the AP-CDW-SDW states in Figure \ref{Fig12}b. The hole density is always maximum at the domain wall with the vanishing spin moment. However, if there is no SDW, such as the AP-CDW stripe in Figure \ref{Fig12}c, then the hole density is maximum at the domain wall with the vanishing pair field. This finding is different from previous work without including the renormalized chemical potential[37].  

\vspace{2mm}

\begin{figure}[ht]
\centering
\includegraphics[width=\linewidth]{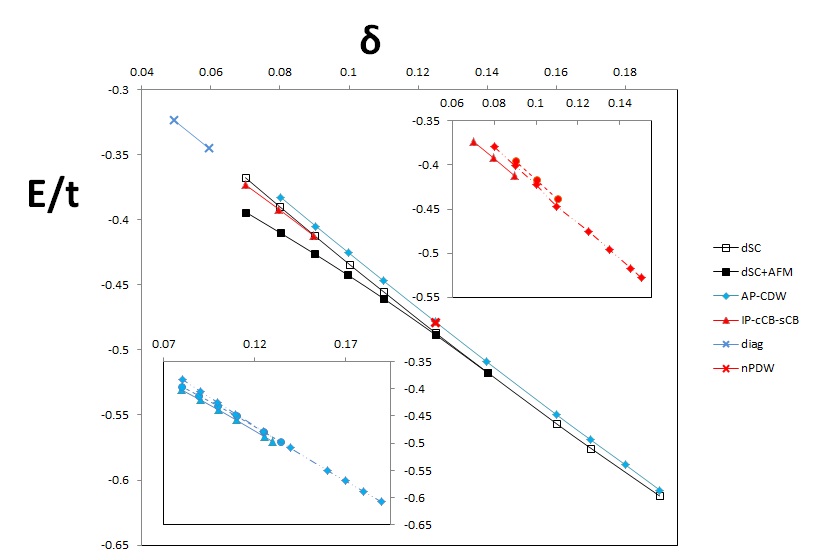}
\caption{}
\label{Fig22}
\end{figure}

Figure \ref{Fig22} shows energies as a function of hole concentration for  all the states listed in Table \ref{tab:example}. The three unidirectional states are shown in the lower inset with blue triangles, circles, and diamonds representing IP-CDW-SDW, AP-CDW-SDW, and AP-CDW, respectively.The three CB states are shown in the upper inset with red triangles, circles and diamonds representing IP-cCB-sCB, AP-cCB-sCB, and AP-cCB, respectively. Unless specifically mentioned, we only report site-centred results. Bond-centred solutions have essentially the same energies. The same results for the three CDW states were also reported in Ref.[30] at a $1/8$  hole concentration. These mean-field GWA results are quite consistent with the numerical Monte Carlo result [29], which revealed that the uniform state has the lowest energy, followed by the in-phase stripe, and that the energy of the anti-phase stripe is slightly above that of both of them. However, the small energy differences are insignificant compared to the result of iPEPS[43], which showed the same ordering of states but with essentially degenerate energies.

\vspace{2mm}

At approximately 12$\%$ doping in Figure \ref{Fig22}, the spin moment becomes smaller, and the uniform dSC-AFM state merges into the dSC state. The difference from the original work of Ogata and Himeda[32,33], in which the spin moment vanished at 10$\%$ doping, is due to the simplified Gutzwiller factors used in Eq. (4). All the magnetic states, such as SDW and sCB, vanish at approximately 12$\%$ doping. The most surprising and important result shown in Figure \ref{Fig22} is that in addition to the uniform dSC state, the AP-CDW state is most stable for a large doping range, from  0.08 to 0.18. The AP-cCB state also extends a little bit beyond the antiferromagnetic region. We only find the diagonal stripe state up to 6$\%$ doping. Another pattern that seems to be limited to small doping is IP-cCB-sCB, which is  only found at doping less than 0.1. The general locations of these CB states in Figure \ref{Fig22} are consistent with experimental observations that CB are seen more often at low doping [12,13]. Because the Gutzwiller factor $g^t_{i,j\sigma}$ in Eq. (4) is proportional to the hole density at the site, we expect the kinetic energy to be maximum at the domain wall (Figure \ref{Fig12}c), as shown in Table \ref{tab:example2}. Table \ref{tab:example2} lists the values of hole density, the magnitude of the pairing order parameter and the kinetic energy $K$ at each site, which are calculated by averaging the four nearest neighbour hopping amplitudes for AP-CDW at a $1/8$ hole concentration. The kinetic energy and pairing order are calculated from the variational parameters  $\chi_{ij\sigma}^v$ and $\Delta_{ij\sigma}^v$ respectively, by using Eq. (S9) in SM. Similar tables for other stripes and CB patterns are presented in Tables 3 and 4 in the SM.

\vspace{2mm}

The red cross in Figure \ref{Fig22} at the $1/8$ hole concentration is the energy of a solution as a result of relaxing the requirement to have a commensurate $4a_0$  period for  the AP-CDW state.  To alleviate the difficulty of considering incommensurate solutions in a finite lattice calculation, we allow the state to have more than one single modulation period. In Figure \ref{Fig32}, the hole density, listed as the red numbers below the pattern, along with the magnitude of the pairing order parameter for both x and y bonds, listed in the top and bottom rows,  are plotted along the direction of the modulation for a complex bond-centred stripe of length 16$a_0$. It is very similar to the AP-CDW state. However, there is a remaining net constant d-wave pairing, with the system average $\Delta_x=-0.0056$ and $\Delta_y=0.0057$.  This mixture of the AP-CDW stripe with a small constant uniform pairing will produce a d-wave nodal-like LDOS in addition to a PDW; hence, we have a nodal PDW or nPDW. There are several important results associated with the nPDW.  Figure \ref{Fig32} shows that the hole density is indeed maximum at the domain walls near sites 4,7,10 and 13. The maximum amplitude of pairing order $\Delta$ is about 0.03, which is roughly the same as adding the net pairing amplitude to that of the AP-CDW stripe in Table 2. It is most gratifying to observe that the d-wave pairing is globally maintained, although we have no way of controlling it during the iteration, with variables changing independently on each site.  Contrary to the pure AP-CDW state without a net pairing, this state has a d-wave nodal spectrum at low energy, hence a nodal-like LDOS.  In Figure \ref{Fig42}a, the LDOS of this stripe at 8 sites is plotted as a function of energy.  The positions of these 8 sites are indicated in the inset of Figure \ref{Fig42}a. The detailed LDOS at low energy is shown in Figure \ref{Fig42}b. The large spatial variation of LDOS at high energies but always with a d-wave node  near zero energy is quite consistent with the STM results in Ref.[3]. We have obtained this result by using a lattice of 16x16 supercells; please see the SM for details. 

\vspace{2mm}

A special feature of all these charge-ordered states is the large variation of the Gutzwiller factors from site to site. The values could change between nearest neighbours by a factor of 2 to 3. This  unique property of strong correlated systems originates from the dependence on local hole density in the Gutzwiller factor, which is  $g^t_{i}=\sqrt{\frac{2\delta_i}{1+\delta_i}}$, when we do not consider magnetic moments. This  dependence on ${\delta_i}$ is the consequence of being a  Mott insulator when there are no doped holes. A slight variation of the hole density $\delta_i$ will cause a large change in $g^t_{i}$; in fact, ${\partial g^t_{i}}/{\partial \delta_i}$ is proportional to ${g^t_{i}}/{\delta_i} \sim 1/\sqrt{\delta_i}$. This factor dominates in the renormalized local chemical potential defined in Eq. (S6)  when hole concentration is small. Thus, $g^t_{i}$ is no longer a purely passive renormalization factor; now, it could alter the local chemical potential greatly and induce non-uniform charge orders. Although the factor associated with spin, $g^{s,xy}_{i}$ in Eq. (4), is smaller, it also contributes to the local chemical potential. The strong susceptibility to the variation of local hole density makes a uniform state  unstable amidst inherent or extrinsic charge fluctuations. This effect is clearly more prominent in the lightly hole-doped regime, as demonstrated by the greater variety of charge-ordered states in the underdoped regime in Figure  \ref{Fig22}. Another important effect of the Gutzwiller factor is that it introduces nonlinearity into the Bogoliubov-deGenne(BdG) equations(Eq. (S4)-(S6)), which can produce quite unexpected solutions.  

\vspace{2mm}

\subsection*{Bond Order}

So far, we have only discussed the pair field, hole density and spin moment; now, we shall consider more carefully the bond order $K_{ij}=\frac{1}{2}\sum_{\sigma}\langle c^\dagger_{i\sigma}c_{j\sigma} \rangle+\langle c^\dagger_{j\sigma}c_{i\sigma} \rangle$. The value of one-half in front of the summation is for averaging because there are two hopping terms for each bond. Now, it can be calculated by using the BdG solution and the Gutzwiller factor, i.e., $K_{ij}=\frac{1}{2}\sum_{\sigma}g^t_{ij\sigma}\chi_{ij\sigma}^v+g^t_{ji\sigma}\chi_{ji\sigma}^v$. Following the definition of bond order by Sachdev and collaborators [21,22,40] and Fujita \textsl{et al.}[16], by associating $K_{i;i+\hat{x}}\sim\rho(\textbf{r}_{O_x})$, the tunneling current measured at the x-bond oxygen site can be obtained, similarly for the y-bond oxygen. The Fourier transform of these two quantities gives us the intra-unit-cell form factor. The Fourier transform of the
AP-CDW state with a hole concentration of $1/8$ is schematically shown in Figure \ref{Fig52}a. The size of the dot represents
the magnitude; red (blue) represents a positive (negative) value. Because this is a $4a_0$ stripe, in addition to values at $Q=(0,0)$ and reciprocal lattice vectors denoted by the ``+" sign, the modulation wave vector is  $(\pm\pi/2a_0,0)$, and the vectors are shifted by the reciprocal lattice vectors. The peaks at $(\pm\pi/2a_0,0)$ are determined by $A_{S'}$, while those at $(\pm3\pi/2a_0,0)$ and $(\pm \pi/2a_0,\pm 2\pi a_0)$  are determined by  $A_{D}$. The ratio of $A_{D}$ to $A_{S'}$, or $d/s'$, is approximately 7.5 in this case. This ratio is quite special for the AP-CDW state. For the IP-CDW-SDW stripe, the ratio is actually less than one. The schematic plots of the Fourier transform of IP-CDW-SDW and AP-CDW-SDW stripes are shown in Figure 8a and 8b in the SM, respectively. For the AP-CDW-SDW stripe, $d/s'$ is approximately $1.2$. The Fourier transform of the bond orders of the AP-cCB pattern is similar to that of AP-CDW with a dominant d-form factor, as discussed in the SM.

\vspace{3mm}

\begin{figure}[ht]
\centering
\includegraphics[width=\linewidth]{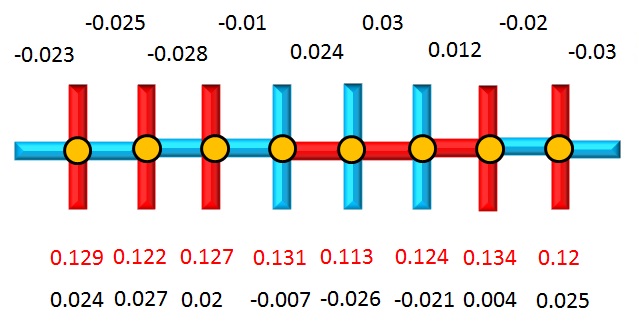}
\caption{}
\label{Fig32}
\end{figure}

\begin{figure}[ht]
\centering
\includegraphics[width=\linewidth]{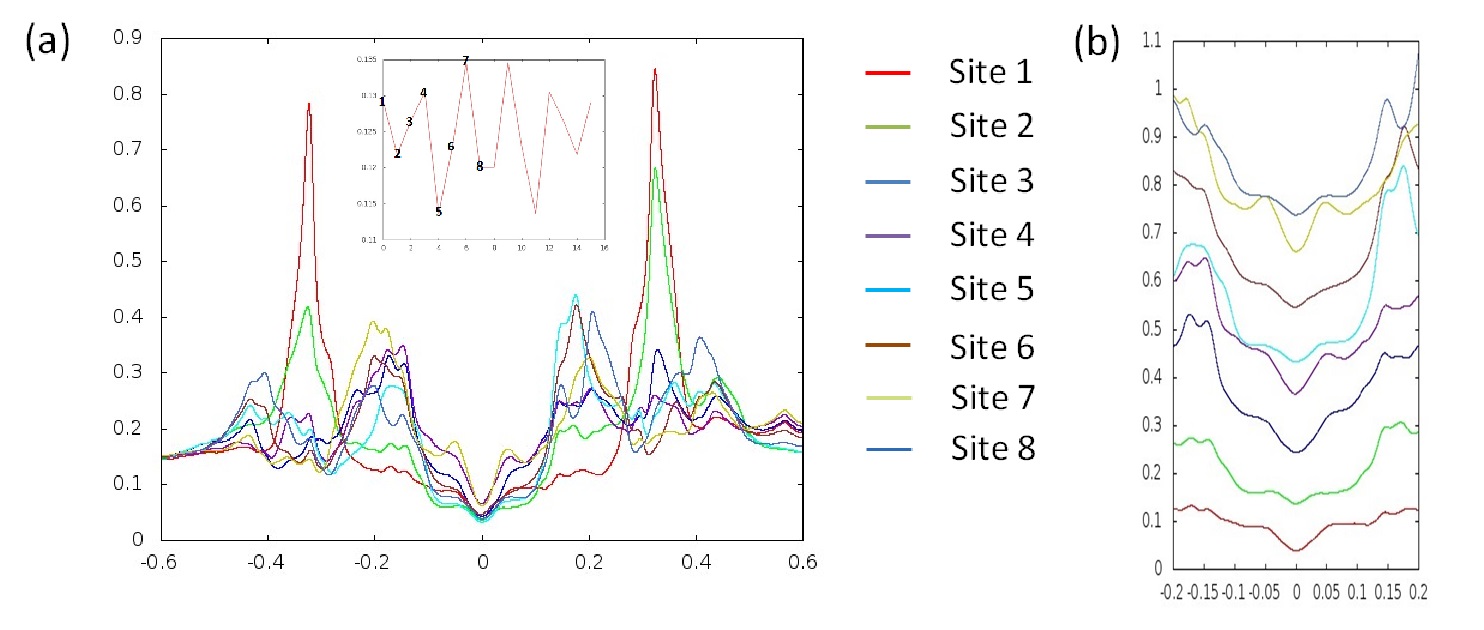}
\caption{}
\label{Fig42}
\end{figure}

\begin{figure}[ht]
\centering
\includegraphics[width=\linewidth]{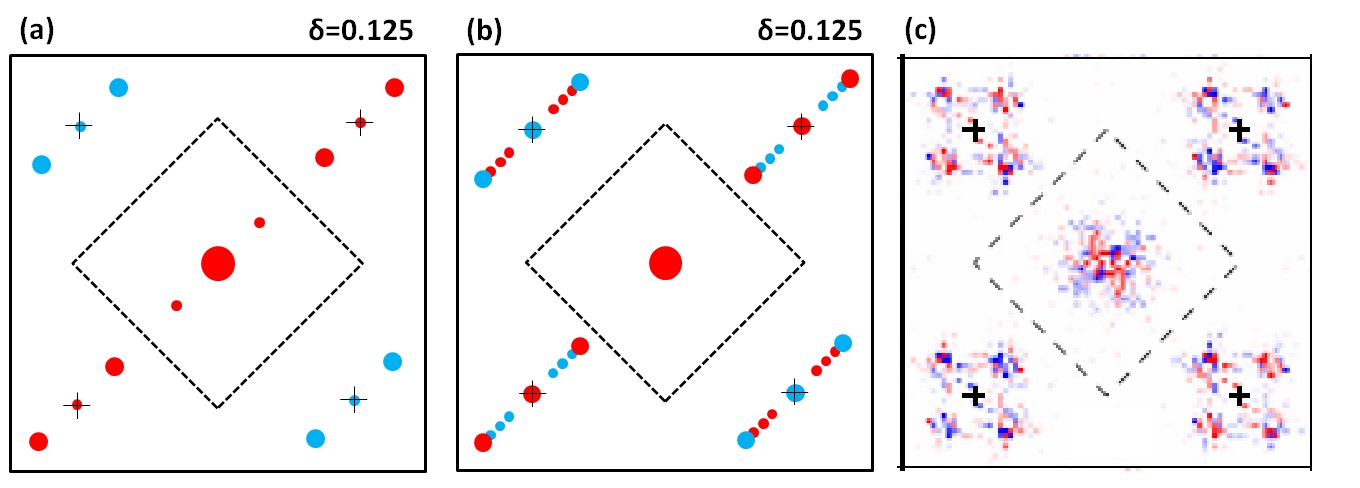}
\caption{}
\label{Fig52}
\end{figure}

The nPDW stripe shown in Figure \ref{Fig32} also has a large $d$-form factor with almost zero $s'$. The  Fourier transform of its bond order is schematically shown in Figure \ref{Fig52}b. The size of the dot scales with the magnitude of the $d$-form factors, and red (blue) represents a positive (negative) value. The wave vector with a large amplitude is at $5\pi/8a_0$ or its period is approximately $3.2a_0$. This length is close to the separation between the domain walls of the pair field shown in Figure \ref{Fig32}. The presence of  smaller peaks at several wave vectors shows a mixture of different periods in the stripe. This result is expected if we add a constant pairing order to the AP-CDW stripe.

\vspace{3mm}
Figure \ref{Fig52}c is copied from  Figure 3G of the STS work of Fujita \textsl{et al.}[16]. It shows the sum of real part of Fourier transform values of tunneling currents measured at $O_x$ and $O_y$ sites.  Just like Figure \ref{Fig52}a and Figure \ref{Fig52}b, The value at $(\pm3\pi/2a_0,0)$ is larger than that at $(\pm\pi/2a_0,0)$ and both have the same sign but opposite sign with respect to $(\pm \pi/2a_0,\pm 2\pi a_0)$.  In their sample there are two domains with density modulation in  x and y directions, respectively. 

\vspace{2mm}

\begin{figure}[ht]
\centering
\includegraphics[width=\linewidth]{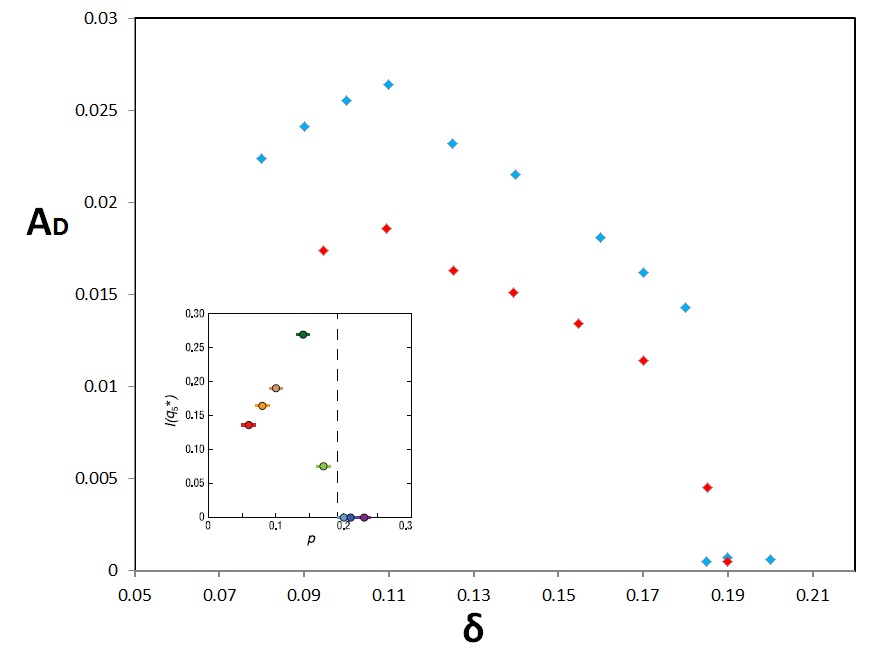}
\caption{}
\label{Fig62}
\end{figure}

Another interesting result regarding the AP-CDW stripe is that its $d$-form factor actually vanishes at an approximately 19$\%$ hole concentration, as shown in Figure \ref{Fig62} for both site-centred (blue dots) and bond-centred (red dots) solutions. We cannot find the AP-CDW solution beyond 18$\%$ doping. This outcome is in excellent agreement with the  results reported by Fujita \textsl{et al.}[17] in their Figure 3G which is copied as the inset of Figure \ref{Fig62}. They measured the doping dependence of intensity of the modulation wave vector near $(\pm3\pi/2a_0,0)$, which is associated with the density wave. The density wave  disappeares at approximately 19$\%$ doping. Moreover, this 19$\%$ hole concentration is conspicuously close to the so-called quantum critical point[46]. We shall study this issue more in future work.

\section*{Conclusion}

The results reported above are all based upon the well-established renormalized mean-field theory[45] and GWA[44] for a well-studied $t-J$ model. Although they do not provide extremely accurate numbers, as many sophisticated numerical methods do, our results show that they do capture the most important physics of the strong correlation.  This strong correlation provides a site-dependent Gutzwiller renormalization that produces many exotic solutions of PDW stripes and/or CBs intertwined with modulations of charge density and/or spin density.  These results show quantitative agreement with some of the key experiments[3,12,13]. Because site-renormalization is extremely  local, the effect of the Fermi surface or wave vectors $k_F$ is absent. Our model does not have the second or third neighbour hopping to provide a Fermi surface with nesting vectors or ``hot spots" [21,40,46]. Thus, in our theory, there are no unique wave vectors for the charge density waves or CBs. Although we have mainly focused on the structures with a period of $4a_0$ so far, our preliminary study also finds  charge-ordered states with periods of $5a_0$ and even $3a_0$. States with a longer period should be possible, and they could also have degenerate energies[34,43]. If we allow a pattern with multiple periods, such as the nPDW stripe shown in Figures \ref{Fig32} and \ref{Fig52}b, we could have states with fractional or incommensurate periods.  A detail study of all these will be conducted in the future, as well as a study of the effect of having values of $J/t$ away from $0.3$.     

\vspace{2mm}

An important consequence of having all these charge-ordered states originating from the same Hamiltonian and physics is that these states are not the usual ``competing states" we are familiar with. They do not stay in a  deep local minima in the energy landscape. They are actually quite fragile and can easily evolve into each other, as we have already demonstrated with the nPDW stripe, which evolved from a mixture of AP-CDW and an uniform d-SC state. Other examples of the mixture of stripes listed in Table \ref{tab:example} can be easily constructed. For real cuprates, there are many other interactions in addition to our $t$ and $J$ that will alter the preferences of these states. For example, a weak electron lattice interaction could make the IP-CDW-SDW stripe much more stable against the dSC-AFM state[36].  Including special Fermi surface features could also enhance CDW for certain periods. However, none of these interactions are as important and necessary as the site renormalization due to strong Mott physics to produce these charge-ordered states. The effect of finite temperature will certainly bring in the entanglement of these states and much more complicated phenomena, such as pseudogap.  Developing a method for generalising GWA to include the temperature effect remains as a big challenge.

\section*{Methods}

We introduce the $t-J$ Hamiltonian[18] on a square lattice of Cu by using
\begin{equation}
\begin{aligned}
H=-\sum_{\langle i,j\rangle,\sigma}P_{G}t(c^\dagger_{i\sigma}c_{j\sigma}+H.C.)P_{G}+\sum_{\langle i,j\rangle}JS_i \cdot S_j
\end{aligned}
\end{equation}
where nearest neighbour hopping $t$, as our energy unit, is set to 1, and $J$ is set to 0.3. $P_G=\prod_i(1-n_{i\uparrow}n_{i\downarrow})$ is the Gutzwiller projection operator, while $n_{i\sigma}=c^\dagger_{i\sigma}c_{i\sigma}$ stands for the number operator for site i.  Spin $\sigma$ is equal to $\pm$. $S_i$ is the spin one-half operator at site i. The Fermi surface of the uniform state is quite simple, without nesting parts, and does not  intersect with the magnetic Brillouin zone boundary, thus avoiding hot spots. 

\vspace{2mm}

Following the idea of Gutzwiller[44] and work of Himeda and Ogata[32,33], we replace the projection operator($P_G$) with the Gutzwiller renormalization factors. The renormalized Hamiltonian now becomes
\begin{equation}
\begin{aligned}
H=-\sum_{i,j,\sigma}g^t_{ij\sigma}t(c^\dagger_{i\sigma}c_{j\sigma}+H.C.)+\sum_{\langle i,j\rangle}J\Bigg [ g^{s,z}_{ij}S^{s,z}_i S^{s,z}_j+g^{s,xy}_{ij}\Bigg(\frac{S^{+}_i S^{-}_j+S^{-}_i S^{+}_j}{2}\Bigg)\Bigg] \\
\end{aligned}
\end{equation}
where $g^t_{ij\sigma}, g^{s,z}_{ij}$, and $g^{s,xy}_{ij}$ are the Gutzwiller factors, which are dependent on the values of local AF moment $m_i^v$,  pair field $\Delta_{ij\sigma}^v$, bond order $\chi_{ij\sigma}^v$, and hole density $\delta_i$: 
\begin{equation}
\begin{aligned}
&m_i^v=\langle\Psi_0 | S^z_i | \Psi_0 \rangle\\
&\Delta_{ij\sigma}^v=\sigma \langle\Psi_0 | c_{i\sigma}c_{j\bar{\sigma}} | \Psi_0 \rangle\\
&\chi_{ij\sigma}^v=\langle\Psi_0 | c^\dagger_{i\sigma}c_{j\sigma} |\Psi_0 \rangle\\
&\delta_i=1-\langle\Psi_0 | n_i |\Psi_0 \rangle
\end{aligned}
\end{equation}
where $| \Psi_0 \rangle$ is the unprojected wavefunction. The superscript $v$ is used to denote that these quantities are different from the real physical quantities for comparison with the experiments. Their relationship is given in Eq. (S9).  As for the Gutzwiller factors, we follow the work of Yang \text{et al.}[30]; they used a slightly simplified version of Ogata and Himeda[32,33], which was also used by Christensen \textsl{et al.}[34]. The factors are given as
\begin{equation}
\begin{aligned}
&g^t_{ij\sigma}=g^t_{i\sigma}g^t_{j\sigma}\\
&g^t_{i\sigma}=\sqrt{\frac{2\delta_i(1-\delta_i)}{1-\delta_i^2+4(m_i^v)^2}\frac{1+\delta_i+\sigma2m_i^v}{1+\delta_i-\sigma2m_i^v}} \\
&g^{s,xy}_{ij}=g^{s,xy}_i g^{s,xy}_j\\
&g^{s,xy}_i=\frac{2(1-\delta_i)}{1-\delta_i^2+4(m_i^v)^2}\\
&g^{s,z}_{ij}=g^{s,xy}_{ij} \frac{2((\bar{\Delta}^v_{ij})^2+(\bar{\chi}^v_{ij})^2)-4m_i^vm_j^vX^2_{ij}}{2((\bar{\Delta}^v_{ij})^2+(\bar{\chi}^v_{ij})^2)-4m_i^vm_j^v}\\
&X_{ij}=1+\frac{12(1-\delta_i)(1-\delta_j)((\bar{\Delta}^v_{ij})^2+(\bar{\chi}^v_{ij})^2)}{\sqrt{(1-\delta_i^2+4(m_i^v)^2)(1-\delta_j^2+4(m_j^v)^2)}}
\end{aligned}
\end{equation}
where $\bar{\Delta}^v_{ij}=\sum_{\sigma}\Delta_{ij\sigma}^v/2$ and $\bar{\chi}^v_{ij}=\sum_{\sigma}\chi_{ij\sigma}^v/2$. In the presence of antiferromagnetism, $\Delta_{ij\uparrow}^v\ne \Delta_{ij\downarrow}^v$. The derivation of the mean-field self-consistent equations is described in the SM.

\section*{Reference}
\begin{enumerate}
\itemsep=-5pt
\item Yamada, K. \textsl{et al}. Doping dependence of the spatially modulated dynamical spin correlations and the superconducting-transition temperature in $La_{2-x}Sr_xCuO_4$. \textsl{Phys. Rev. B} \textbf{57}, 6165-6172 (1998).

\item Abbamonte P. \textsl{et al}. Spatially modulated `Mottness' in $La_{2-x}Ba_xCuO_4$. \textsl{Nature Physics} \textbf{1}. 155-158 (2005).

\item Kohsaka, Y. \textsl{et al}. An intrinsic bond-centered electronic glass with unidirectional domains in underdoped cuprates. \textsl{Science} \textbf{315}, 1380-1385 (2007).

\item Parker, C. V. \textsl{et al}. Fluctuating stripes at the onset of the pseudogap in the high-$T_c$ superconductor $Bi_2Sr_2CaCu_2O_{8+x}$. \textsl{Nature} \textbf{468}, 677-680 (2010).

\item Wu, T. \textsl{et al}. Magnetic-field-induced charge-stripe order in the high-temperature superconductor $YBa_2Cu_3O_y$. \textsl{Nature} \textbf{477}, 191-194 (2011).

\item Ghiringhelli, G. \textsl{et al}. Long-range incommensurate charge fluctuations in $(Y,Nd)Ba_2Cu_3O_{6+x}$. \textsl{Science} \textbf{337}, 821-825 (2012).

\item Comin, R. \textsl{et al}. Charge order driven by Fermi-arc instability in $Bi_2Sr_{2-x}La_xCuO_{6+\delta}$. \textsl{Science} \textbf{343}, 390-392 (2014).

\item da Silva Neto, E. H. \textsl{et al}. Ubiquitous Interplay Between Charge Ordering and High-Temperature Superconductivity in Cuprates. \textsl{Science} \textbf{343}, 393-396 (2014).

\item Hashimoto, M. \textsl{et al}. Direct observation of bulk charge modulations in optimally doped $Bi_{1.5}Pb_{0.6}Sr_{1.54}CaCu_2O_{8+\delta}$. \textsl{Phys. Rev. B} \textbf{89}, 220511(R) (2014).

\item Blanco-Canosa, S. \textsl{et al}. Resonant x-ray scattering study of charge-density wave correlations in $YBa_2Cu_3O_{6+x}$. \textsl{Phys. Rev. B} \textbf{90}, 054513 (2014).

\item da Silva Neto, E. H. \textsl{et al}. Charge ordering in the electron-doped superconductor $Nd_{2-x}Ce_xCuO_4$. \textsl{Science} \textbf{347}, 282-285 (2015).

\item Wise, W. D. \textsl{et al}. Charge-density-wave origin of cuprate checkerboard visualized by scanning tunneling microscopy. \textsl{Nature Physics} \textbf{4}, 696-699 (2008).

\item Hanaguri, T. \textsl{et al}. A `checkerboard' electronic crystal state in lightly hole-doped $Ca_{2-x}Na_xCuO_2Cl_2$. \textsl{Nature} \textbf{430}, 1001-1005 (2004).

\item Comin, R. \textsl{et al}. Symmetry of charge order in cuprates. \textsl{Nature Materials} \textbf{14}, 796-800 (2015).

\item Achkar, A. J. \textsl{et al}. Orbital symmetry of charge density wave order in $La_{1.875}Ba_{0.125}CuO_4$ and $YBa_2Cu_3O_{6.67}$. \textsl{arXiv}:1409.6787 (2014).

\item Fujuta, K. \textsl{et al}. Direct phase-sensitive identification of a d-form factor density wave in underdoped cuprates. \textsl{PNAS.} \textbf{111} 30, E3026-E3032 (2014).

\item Fujita, K. \textsl{et al}. Simultaneous transitions in cuprate momentum-space topology and electronic symmetry breaking. \textsl{Science} \textbf{344}, 612-616 (2014).

\item Anderson, P. W.. The Resonating Valence Bond State in $La_2CuO_4$ and superconductivity. \textsl{Science} \textbf{235}, 1196-1198 (1987).

\item Hashimoto, M., Vishik, I., He, R., Devereaux, T., $\&$ Shen, Z.. Energy gaps in high-transition-temperature cuprate superconductors. \textsl{Nature Physics} \textbf{10}, 483-495 (2014).

\item Zhang, F. C. $\&$ Rice, T. M.. Effective Hamiltonian for the superconducting Cu oxides. \textsl{Phys. Rev. B} \textbf{37}, 3759-3761 (1988).

\item Metlitski, M. $\&$ Sachdev, S.. Instabilities near the onset of spin density wave order in metals. \textsl{New J. Phys.} \textbf{12}, 105007 (2010).

\item Metlitski, M. $\&$ Sachdev, S.. Quantum phase transitions of metals in two spatial dimensions: II. Spin density wave order. \textsl{Phys. Rev. B} \textbf{82}, 075128 (2010).

\item Allais, A., Bauer, J., $\&$ Sachdev, S.. Bond order instabilities in a correlated two-dimensional metal. \textsl{Phys. Rev. B} \textbf{90}, 155114 (2014).

\item Laughlin, R. B.. Hartree-Fock computation of the high-$T_c$ cuprate phase diagram. \textsl{Phys. Rev. B} \textbf{89}, 035134 (2014).

\item Fradkin, E., Kivelson, S., $\&$ Tranquada, J.. Colloquium: Theory of intertwined orders in high temperature superconductors. \textsl{Rev. Mod. Phys.} \textbf{87}, 457-482 (2015).

\item Berg, E., Fradkin, E., Kivelson, S., $\&$ Tranquada, J.. Striped superconductors: how spin, charge and superconducting orders intertwine in the cuprates. \textsl{New J. Phys.} \textbf{11}, 115004 (2009).

\item Loder, F., Graser, S., Kampf, A., $\&$ Kopp, T.. Mean-field pairing theory for the charge-stripe phase of high-temperature cuprate superconductors. \textsl{Phys. Rev. Lett.} \textbf{107}, 187001 (2011).

\item Lee, P. A.. Amperean pairing and the pseudogap phase of cuprate superconductors. \textsl{Phys. Rev. X} \textbf{4}, 031017 (2014).

\item Chou, C., Fukushima, N., $\&$ Lee, T.. Cluster-glass wave function in the two-dimensional extended $t-J$ model. \textsl{Phys. Rev. B} \textbf{78}, 134530 (2008).

\item Yang, K., Chen, W., Rice, T. M., Sigrist, M. $\&$ Zhang, F. C.. Nature of stripes in the generalized $t-J$ model applied to the cuprate superconductors. \textsl{New J. Phys.} \textbf{11}, 055053 (2009).

\item Himeda, A., Kato, T., $\&$ Ogata, M.. Stripe States with Spatially Oscillating d-Wave Superconductivity in the Two-Dimensional $t-t'-J$ Model. \textsl{Phys. Rev. Lett.} \textbf{88}, 117001 (2002).

\item Himeda, A. $\&$ Ogata, M.. Coexistence of $d_{x^2-y^2}$ superconductivity and antiferromagnetism in the two-dimensional $t-J$ model and numerical estimation of Gutzwiller factors. \textsl{Phys. Rev. B} \textbf{60}, R9935-R9938 (1999).

\item Ogata, M. $\&$ Himeda, A.. Superconductivity and antiferromagnetism in an extended Gutzwiller approximation for $t-J$ model: effect of double-occupancy exclusion. \textsl{J. Phys. Soc. Japan} \textbf{72}, 374-391 (2003).

\item Christensen, R. B., Hirschfeld, P. J., $\&$ Anderson, B. M.. Two routes to magnetic order by disorder in underdoped cuprates. \textsl{Phys. Rev. B} \textbf{84}, 184511 (2011).

\item Chou, C. $\&$ Lee, T.. Inhomogeneous state of the extended $t-J$ model on a square lattice: A variational Monte Carlo and Gutzwiller approximation study. \textsl{Phys. Rev. B} \textbf{85}, 104511 (2012).

\item Chou, C. $\&$ Lee, T. Mechanism of formation of half-doped stripes in underdoped cuprates. \textsl{Phys. Rev. B} \textbf{81}, 060503 (2010).

\item Poilblanc, D.. Stability of inhomogeneous superstructures from renormalized mean-field theory of the $t-J$ model. \textsl{Phys. Rev. B} \textbf{72}, 060508 (2005).

\item White, S. $\&$ Scalapino, D. J.. Density matrix renormalization group study of the striped phase in the 2D $t-J$ model. \textsl{Phys. Rev. Lett.} \textbf{80}, 1272-1275 (1998).

\item White, S. $\&$ Scalapino, D. J.. Pairing on striped $t-t'-J$ lattices. \textsl{Phys. Rev. B} \textbf{79}, 220504 (2009).

\item Sachdev, S. $\&$ La Placa, R.. Bond order in two-dimensional metals with antiferromagnetic exchange interactions. \textsl{Phys. Rev. Lett.} \textbf{111}, 027202 (2013).

\item Davis, J. C. $\&$ Lee, D.. Concepts relating magnetic interactions, intertwined electronic orders, and strongly correlated superconductivity. \textsl{PNAS.} \textbf{110}, 17623-17630 (2013).

\item Wang, Y. $\&$ Chubukov, A.. Charge-density-wave order with momentum $(2Q,0)$ and $(0,2Q)$ within the spin-fermion model: Continuous and discrete symmetry breaking, preemptive composite order, and relation to pseudogap in hole-doped cuprates. \textsl{Phys. Rev. B} \textbf{90}, 035149 (2014).

\item Corboz, P., Rice, T. M., $\&$ Troyer, M.. Competing states in the $t-J$ model: uniform d-wave state versus stripe state. \textsl{Phys. Rev. Lett.} \textbf{113}, 046402 (2014).

\item Gutzwiller, M.. Effect of correlation on the ferromagnetism of transition metals. \textsl{Phys. Rev. Lett.} \textbf{10}, 159-162 (1963).

\item Zhang, F. C., Gros, C., Rice, T. M., $\&$ Shiba, H.. A renormalised Hamiltonian approach to a resonant valence bond wavefunction. \textsl{Supercond. Sci. Technol.} \textbf{1}, 36-46 (1988).

\item Efetov, K. B., Meier, H., $\&$ P\'{e}pin, C.. Pseudogap state near a quantum critical point. \textsl{Nature Physics} \textbf{9}, 442-446 (2013).

\end{enumerate}

\section*{Acknowledgements}

We acknowledge and thank T. M. Rice, S. A. Kivelson, and D.H. Lee for helpful conversations and communications. We are particular in debt to  Mohammad H. Hamidian for sharing his slides and insights. This work was partially supported by Taiwan Ministry of Science and Technology with Grant No.101-2112-M-001-026-MY3 and calculation was supported by the National Center for High Performance Computing in Taiwan.

\section*{Author contributions statement}

TK.L. conceived the original idea. W.T. and TK.L. provided the theoretical understanding and wrote the paper together.

\section*{Additional information}
\subsection*{Competing financial interests}The authors declare no competing financial interests.

\section*{Figure Legends}

\subsection*{Figure 1}

Schematic illustration of modulations for stripe like patterns: (a)IP-CDW-SDW (b)AP-CDW-SDW (c)AP-CDW respectively. Size of the circle represents the hole density. The width of the bond around each site represents the amplitude of pairing $\Delta$($\Delta=\sum_{\sigma}\Delta_{\sigma}$) and sign is positive(negative) for red(cyan). The size of black arrows represents the spin moment. The average hole density is about 0.1.

\subsection*{Figure 2}

Energy per site as a function of hole concentration. Six states are shown in the main figure with notations defined in Table \ref{tab:example2}.   The lower(upper) inset is  for stripe(CB) patterns. Blue triangles, circles, and diamonds  are for IP-CDW-SDW, AP-CDW-SDW, and AP-CDW respectively. And red triangles, circles and diamonds are for IP-cCB-sCB, AP-cCB-sCB, and AP-cCB respectively.

\subsection*{Figure 3}

Schematic illustration of modulations for nPDW stripe. The numbers in red denote the hole dnesity at each site while the numbers in black below them represent the pairing amplitude in y direction. The rest numbers above the figure stand for the pairing amplitude in x direction. Here our pairing amplitudes denote $(\langle c_{i\uparrow}c_{j\downarrow} \rangle)$. Note that in this figure neither the size of circles nor the width of bonds represent amplitudes. The hole concentration is 0.125.

\subsection*{Figure 4}

(a) LDOS at 8 sites plotted from energy  0.6t to -0.6t. The inset shows hole density along the modulation direction of the nPDW stripe and (b) from 0.2t to -0.2t but shifted vertically for clarity.

\subsection*{Figure 5}

Schematic illustration of the Fourier transform of bond orders of (a) AP-CDW state and (b)  the nPDW stripe in a lattice of $16a_0*16a_0$. "+" signs are at the four reciprocal 
lattice vectors $(\pm2\pi/a_0,0)$ and $(0,\pm2\pi/a_0)$ and their nearby medium size dots are shifted from them by $(\pm\pi/2a_0,0)$.  The large dot at center is $Q=(0,0)$ and has two red small dots nearby at $(\pm\pi/2a_0,0)$.
The inner dotted square is the boundary of first Brillouin zone. (c) is copied from  Figure 3G of the STS work of Fujita \textsl{et al.}[16]. It shows the sum of real part of Fourier transform values of tunneling currents measured at $O_x$ and $O_y$ sites. Unlike (a) and (b) that only has one domain of  density modulation in the x direction, this sample has two domains with both  x and y direction modulations. 

\subsection*{Figure 6}

Magnitude of the d-form factor for the AP-CDW stripe as a function of doped hole concentration. Blue dots are for site-centered AP-CDW stripe and red ones for bond-centered AP-CDW. The inset is   copied from Figure 3G of the STS work of Fujita \textsl{et al.}[17] showing the doping dependence of intensity of the modulation wave vector near $(\pm3\pi/2a_0,0)$, which is associated with the density wave. This modulation vanishes at 19$\%$ hole concentration.

\begin{table}[ht]
\centering
\begin{tabular}{|l|l|l|l|}
\hline
    & pair field & charge modulation & spin modulation\\
\hline
IP-CDW-SDW & in-phase & stripe & yes\\
\hline
AP-CDW-SDW & anti-phase & stripe & yes\\
\hline
AP-CDW & anti-phase & stripe & zero\\
\hline
IP-cCB-sCB & in-phase & checkerboard & yes\\
\hline
AP-cCB-sCB & anti-phase & checkerboard & yes\\
\hline
AP-cCB & anti-phase & checkerboard & zero\\
\hline
dSC & uniform & uniform & zero\\
\hline
dSC-AFM & uniform & uniform & uniform\\
\hline
diag & in-phase & stripe along (1,1) & yes\\
\hline
\end{tabular}
\caption{\label{tab:example}Definition of various nearly degenerate states with respect to  the intertwined orders: pair field, charge density, and spin moment. Besides the two uniform solutions, d-wave superconducting (dSC) state and coexistent antiferromagnetic (dSC-AFM) state, all the states to be considered in this paper, unless specifically mentioned, have  modulation period  $4a_0$ for charge density and bond order.  IP(AP) means the pair field is in-phase with period $4a_0$ (anti-phase with period $8a_0$). IP has a net pairing order and AP has none. SDW is the spin density wave with period $8a_0$. sCB (cCB) denotes the checkerboard pattern of spin (charge) and diag means the diagonal stripe which has in-phase pair field and  spin modulation.}
\end{table}

\begin{table}[ht]
\centering
\begin{tabular}{|l|l|l|l|l|}
\hline
 site number   & 1 & 2 & 3 & 4\\
\hline
$\delta_i$ & 0.1315 & 0.1256 & 0.1168 & 0.1256\\
\hline
$\Delta_i$ & 0 & 0.0194 & 0.0247 & 0.0194\\
\hline
$K_i$ & 0.092 & 0.0866 & 0.0799 & 0.0866\\
\hline
$K_{i,i+\hat{y}}$ & 0.1151 & 0.0901 & 0.0625 & 0.0901\\
\hline
$K_{i,i+\hat{x}}$ & 0.0688 & 0.0972 & 0.0972 & 0.0688\\
\hline
\end{tabular}
\caption{\label{tab:example2}Hole density and order parameters at each site for an AP-CDW stripe at 0.125 doping. $\Delta_i$ is the average of pairing order of the four bonds at site $i$. $K_i$ is the average kinetic energy at each site and $K_{i,i+\hat{y}}$ ($K_{i,i+\hat{x}}$) are the bond orders in the y (x) direction.  These parameters are calculated according to Eq. (S9).}
\end{table}

\newpage

%
%
%
%
%
%
%


\begin{center}
\Large\textbf{Supplementary Materials for:}

\Large\textbf{Genesis of charge orders in high temperature superconductors}

\vspace{3mm}

\large{Wei-Lin Tu, Ting-Kuo Lee}
\end{center}


Following the renormalized mean-field theory[1] by using the GWA as in the  works of Yang \textsl{et al.}[2], we derive the formula we used for solving the BdG equations.   
After we replace the projection operator by the Gutzwiller factors and use the mean-field order parameters defined in Eq.(3), the energy of the renormalized Hamiltonian(Eq.(2)) becomes 
\begin{equation}\label{first}
\begin{aligned}
E=\langle\Psi_0 \mid H \mid\Psi_0 \rangle=&-\sum_{i,j,\sigma}g^t_{ij\sigma}t(\chi_{ij\sigma}^v+H.C.)-\sum_{\langle i,j \rangle\sigma}J\Big (\frac{g^{s,z}_{ij}}{4}+\frac{g^{s,xy}_{ij}}{2}\frac{\Delta^{v\ast}_{ij\bar{\sigma}}}{\Delta^{v\ast}_{ij\sigma}}\Big )\Delta^{v\ast}_{ij\sigma}\Delta^v_{ij\sigma}\\
&-\sum_{\langle i,j \rangle\sigma}J\Big (\frac{g^{s,z}_{ij}}{4}+\frac{g^{s,xy}_{ij}}{2}\frac{\chi^{v\ast}_{ij\bar{\sigma}}}{\chi^{v\ast}_{ij\sigma}}\Big )\chi^{v\ast}_{ij\sigma}\chi^v_{ij\sigma}+\sum_{\langle i,j \rangle}g^{s,z}_{ij}Jm^v_im^v_j  
\end{aligned}
\tag{S1}
\end{equation} 

Next we want to minimize the energy under two constraints:$\sum_in_i=N_e$ and $\langle\Psi_0 | \Psi_0 \rangle=1$. Thus our target function to be minimized is
\begin{equation}
W=\langle\Psi_0 | H |\Psi_0 \rangle-\lambda(\langle\Psi_0 | \Psi_0 \rangle-1)-\mu\big(\sum_in_i-N_e\big) 
\tag{S2}
\end{equation} 
The mean-field Hamiltonian now becomes
\begin{equation}
\begin{aligned}
H_{MF}=&\sum_{i,j,\sigma}\frac{\partial W}{\partial \chi^v_{ij\sigma}}c^\dagger_{i\sigma}c_{j\sigma}+H.C.+\sum_{\langle i,j \rangle\sigma}\frac{\partial W}{\partial \Delta^v_{ij\sigma}}\sigma c_{i\sigma}c_{j\bar{\sigma}}+H.C.+\sum_{i,\sigma}\frac{\partial W}{\partial n_{i\sigma}}n_{i\sigma}
\end{aligned}
\tag{S3}
\end{equation} 
Eq.(S3) satisfies the Schr$\ddot{o}$dinger equation $H_{MF}| \Psi_0 \rangle=\lambda| \Psi_0 \rangle$. The three derivatives are defined as 
\begin{equation}
\begin{aligned}
H_{ij\sigma}=\frac{\partial W}{\partial \chi^v_{ij\sigma}}=&-J\Big (\frac{g^{s,z}_{ij}}{4}+\frac{g^{s,xy}_{ij}}{2}\frac{\chi^{v\ast}_{ij\bar{\sigma}}}{\chi^{v\ast}_{ij\sigma}}\Big )\chi^{v\ast}_{ij\sigma}-g^t_{ij\sigma}t+\frac{\partial W}{\partial g^{s,z}_{ij}}\frac{\partial g^{s,z}_{ij}}{\partial \chi^v_{ij\sigma}}\\
\end{aligned}
\tag{S4}
\end{equation} 

\begin{equation}
\begin{aligned}
D^\ast_{ij}=\frac{\partial W}{\partial \Delta^v_{ij\uparrow}}=&-J\Big (\frac{g^{s,z}_{ij}}{4}+\frac{g^{s,xy}_{ij}}{2}\frac{\Delta^{v\ast}_{ij\downarrow}}{\Delta^{v\ast}_{ij\uparrow}}\Big )\Delta^{v\ast}_{ij\uparrow}+\frac{\partial W}{\partial g^{s,z}_{ij}}\frac{\partial g^{s,z}_{ij}}{\partial \Delta^v_{ij\uparrow}}\\
\end{aligned}
\tag{S5}
\end{equation} 
and the effective local chemical potential is defined as
\begin{equation}
\begin{aligned}
\mu_{i}=-\frac{\partial W}{\partial n_{i\sigma}}=&\mu-\frac{1}{2}\sigma\sum_jg^{s,z}_{ij}Jm^v_j-\sum_j\frac{\partial W}{\partial g^{s,xy}_{ij}}\frac{\partial g^{s,xy}_{ij}}{\partial n_{i\sigma}}-\sum_j\frac{\partial W}{\partial g^{s,z}_{ij}}\frac{\partial g^{s,z}_{ij}}{\partial n_{i\sigma}}-\sum_{j\sigma'}\frac{\partial W}{\partial g^t_{ij\sigma'}}\frac{\partial g^t_{ij\sigma'}}{\partial n_{i\sigma}}
\end{aligned}
\tag{S6}
\end{equation} 
The last term ${\partial g^t_{ij\sigma'}}/{\partial n_{i\sigma}}$ in the effective local chemical potential gives the biggest contribution.   This was not included in previous works using GWA to study CB[3,4], and their results have different patterns compared with ours. In addition the energy variation between different charge-ordered states is much larger than our nearly degenerate results.  

\vspace{2mm}

Now $H_{MF}$ can be rewritten as BdG equations,
\begin{equation}
\begin{aligned}
H_{MF}= 
\left(
	c^\dagger_{i\uparrow},c_{i\downarrow}
\right)
\left(
	\begin{array}{cc}
	H_{ij\uparrow} & D_{ij}\\
	D^\ast_{ji} & -H_{ji\downarrow}\\
	\end{array}
\right)
\left(
	\begin{array}{cc}
	c_{j\uparrow} \\
	c^\dagger_{j\downarrow} \\
	\end{array}
\right)
\end{aligned}
\tag{S7}
\end{equation}
We can diagonalize the $H_{MF}$ to obtain equal number of positive and negative eigenvalues with their corresponding eigenvectors $(u^n_i,v^n_i)$. With these eigenvectors, we can determine the order parameters at zero temperature by following equations
\begin{equation}
\begin{aligned}
&n_{i\uparrow}=\langle c^\dagger_{i\uparrow}c_{i\uparrow} \rangle_0=\sum_{n_-} |u^n_i|^2 \\
&n_{i\downarrow}=\langle c^\dagger_{i\downarrow}c_{i\downarrow} \rangle_0=\sum_{n_+} |v^n_i|^2\\
&\Delta^v_{ij\uparrow}=\langle c_{i\uparrow}c_{j\downarrow} \rangle_0=\sum_{n_+} u^n_iv^{n\ast}_j\\
&\Delta^v_{ij\downarrow}=-\langle c_{i\downarrow}c_{j\uparrow} \rangle_0=\sum_{n_+} u^n_jv^{n\ast}_i\\
&\chi^v_{ij\uparrow}=\langle c^\dagger_{i\uparrow}c_{j\uparrow} \rangle_0=\sum_{n_-} u^n_ju^{n\ast}_i\\
&\chi^v_{ij\downarrow}=\langle c^\dagger_{i\downarrow}c_{j\downarrow} \rangle_0=\sum_{n_+} v^n_iv^{n\ast}_j\\
\end{aligned}
\tag{S8}
\end{equation} 
The sum for $n_+$($n_-$) means the set of eigenvectors with positive(negative) energies. An iterative method is used to solve  $H_{MF}$ self-consistently. The convergence is achieved for every order parameter if its value changes less than $10^{-3}$ between successive iterations. After the self-consistency is achieved, we calculate order parameters, their formula are
\begin{equation}
\begin{aligned}
\Delta_i=&\sum_{\sigma}(g^t_{i,\sigma}g^t_{i+\hat{x},\bar{\sigma}} \Delta^v_{i,i+\hat{x},\sigma}+g^t_{i,\sigma}g^t_{i-\hat{x},\bar{\sigma}}  \Delta^v_{i,i-\hat{x},\sigma}-g^t_{i,\sigma}g^t_{i+\hat{y},\bar{\sigma}}  \Delta^v_{i,i+\hat{y},\sigma}-g^t_{i,\sigma}g^t_{i-\hat{y},\bar{\sigma}}  \Delta^v_{i,i-\hat{y},\sigma})/8,\\
m_i =& (\sqrt{g^{s,z}_{i,i+\hat{x}}}+\sqrt{g^{s,z}_{i,i-\hat{x}}}+\sqrt{g^{s,z}_{i,i+\hat{y}}}+\sqrt{g^{s,z}_{i,i-\hat{y}}})m^v_{i}/4 ,\\
K_{i,i+\hat{x}}=&\frac{1}{2}\sum_{\sigma}g_{i,i+\hat{x},\sigma}^t \langle c^\dagger_{i\sigma}c_{i+\hat{x}\sigma}\rangle+g_{i+\hat{x},i,\sigma}^t \langle c^\dagger_{i+\hat{x}\sigma}c_{i\sigma}\rangle,\\
K_{i,i+\hat{y}}=&\frac{1}{2}\sum_{\sigma}g_{i,i+\hat{y},\sigma}^t \langle c^\dagger_{i\sigma}c_{i+\hat{y}\sigma}\rangle+g_{i+\hat{y},i,\sigma}^t \langle c^\dagger_{i+\hat{y}\sigma}c_{i\sigma}\rangle,\\
K_i=&(K_{i,i+\hat{x}}+K_{i,i-\hat{x}}+K_{i,i+\hat{y}}+K_{i,i-\hat{y}} )/4
\end{aligned}
\tag{S9}
\end{equation}

The values for the above quantities was shown in Table 2 of the main text  for a typical AP-CDW stripe. Here we show the values for two other stripes in Table \ref{TabS1} and three CB patterns in Table \ref{TabS2}. 

\vspace{2mm}

\begin{figure}[ht]
\centering
\includegraphics[width=3in]{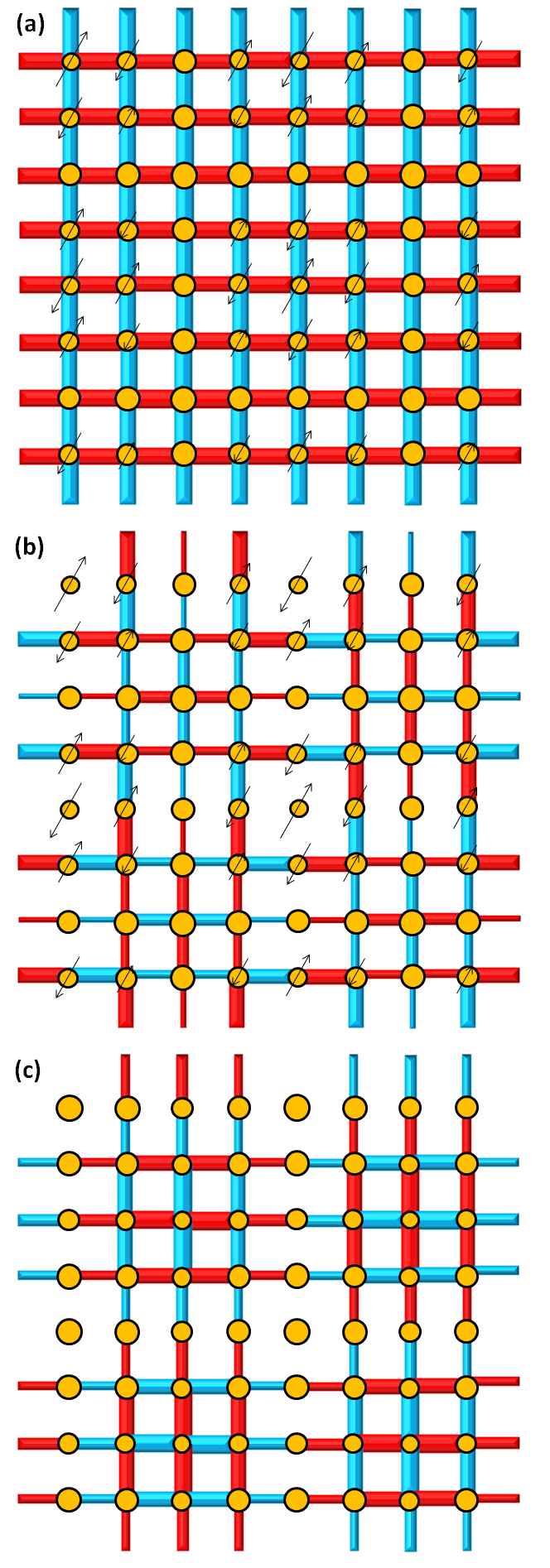}
\caption{Schematic illustration of modulations for CB like patterns: (a)IP-cCB-sCB (b)AP-cCB-sCB (c)AP-cCB respectively. Definitions of all symbols are  same as  Figure 1 of the main text.  The average hole density is 0.1 for AP-cCB-sCB and AP-cCB and 0.09 for IP-cCB-sCB. }
\label{FigS1}
\end{figure}

A schematic illustration of CB  like patterns is shown in Figure \ref{FigS1}. Definitions of symbols are same as Figure 1 in the main text. Again, same as  stripes shown in Figure 1 from the main text, we have the maximum hole density at sites either on the domain walls of AFM or pair field if AFM is absent. The latter part is different from previous results using GWA to study CB[3,4]. 

\vspace{2mm}

We can also examine the symmetry of bond orders, $K_{i,i+\hat{x}}$ and $K_{i,i+\hat{y}}$, as we did in Figure 5 of the main text by examining the Fourier transform. CB in Figure \ref{FigS3} shows clearly it can be thought of as the linear combination of stripes in x and y direction. The small dots inside the dotted square or the first Brillouin zone is proportional to the $s'$ form factor or 
$A_{S'}$ discussed in the main text, and the outside dots are related to the d-form factor or $A_{D}$. Just like AP-CDW stripe, the AP-cCB in Figure \ref{FigS3}c also has a much larger ratio of $d/s'$. For IP-CDW-SDW stripe and IP-cCB-sCB, the ratio is less than one. For AP-CDW-SDW and AP-cCB-sCB, the ratio is about 1.  

\vspace{2mm}

We can also use the BdG solutions to calculate LDOS as shown in Figure 4 of the main text for the nPDW stripe.  Here we use the supercell method[5] to calculate LDOS. Each cell has  $N_x\times N_y$ lattice points and we have $M_c=M_x\times M_y$ cells. We can now reduce the Hamiltonian from $2M_xN_x\times 2M_yN_y$ to $M_x\times M_y$ matrix equations each with dimension $2N_x\times 2N_y$.  LDOS is calculated by the equation
\begin{equation}
\begin{aligned}
\rho_i(E)=&\frac{1}{M_c} \sum_{\textbf{K},n}[(g^t_{i\uparrow})^2 |u^n_i(\textbf{K})|^2\delta(E-E_n(\textbf{K}))+(g^t_{i\downarrow})^2 |v^n_i(\textbf{K})|^2\delta(E+E_n(\textbf{K}))]
\end{aligned}
\tag{S10}
\end{equation}
where $\textbf{K}=2\pi(\frac{n_x}{M_xN_x},\frac{n_y}{M_yN_y})$, $n_x\in[0,M_{x}-1]$ and $n_y\in[0,M_{y}-1]$. Also we replace the delta function by a Lorentzian  function with the width set to be 0.01t in this paper. In Figure 4 of the main text,  we had used $M_x=M_y=N_x=N_y=16$.

\section*{Reference}
\begin{enumerate}
\itemsep=-5pt
\item Zhang, F. C., Gros, C., Rice, T. M., $\&$ Shiba, H.. A renormalised Hamiltonian approach to a resonant valence bond wavefunction. \textsl{Supercond. Sci. Technol.} \textbf{1}, 36 (1988).

\item Yang, K., Chen, W., Rice, T. M., Sigrist, M. $\&$ Zhang, F. C.. Nature of stripes in the generalized $t-J$ model applied to the cuprate superconductors. \textsl{New J. Phys.} \textbf{11}, 055053 (2009).

\item Huang, H., Li, Y., $\&$ Zhang, F. C.. Charge-ordered resonating valence bond states in doped cuprates. \textsl{Phys. Rev. B} \textbf{71}, 184514 (2005).

\item Poilblanc, D.. Stability of inhomogeneous superstructures from renormalized mean-field theory of the $t-J$ model. \textsl{Phys. Rev. B} \textbf{72}, 060508 (2005).

\item Schmid, M., Anderson, B., Kampf, A., $\&$ Hirschfeld, P. J.. d-Wave Superconductivity as a Catalyst for Antiferromagnetism in Underdoped Cuprates. \textsl{New J. Phys.} \textbf{12}, 053043(2010).

\end{enumerate}

\begin{table}[ht]
\centering
\includegraphics[width=\linewidth]{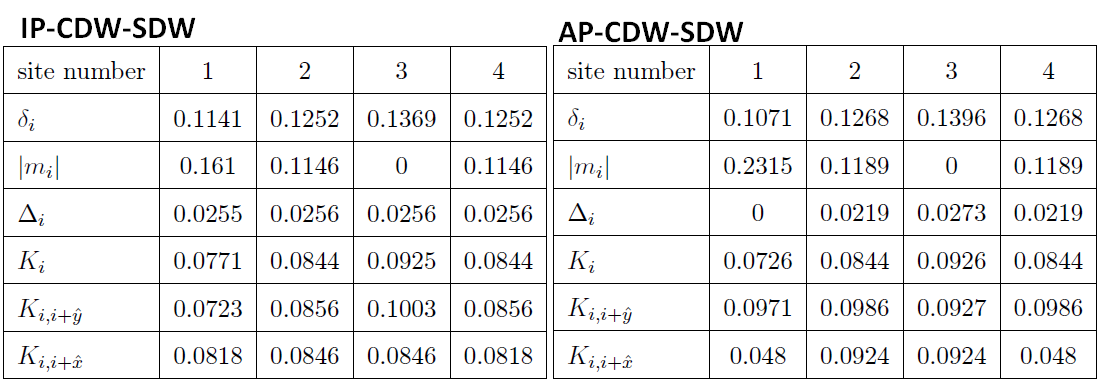}
\caption{Values of several order parameters for IP-CDW-SDW and AP-CDW-SDW stripes at 0.125 doping.}
\label{TabS1}
\end{table}

\begin{table}[ht]
\centering
\includegraphics[width=4.5in]{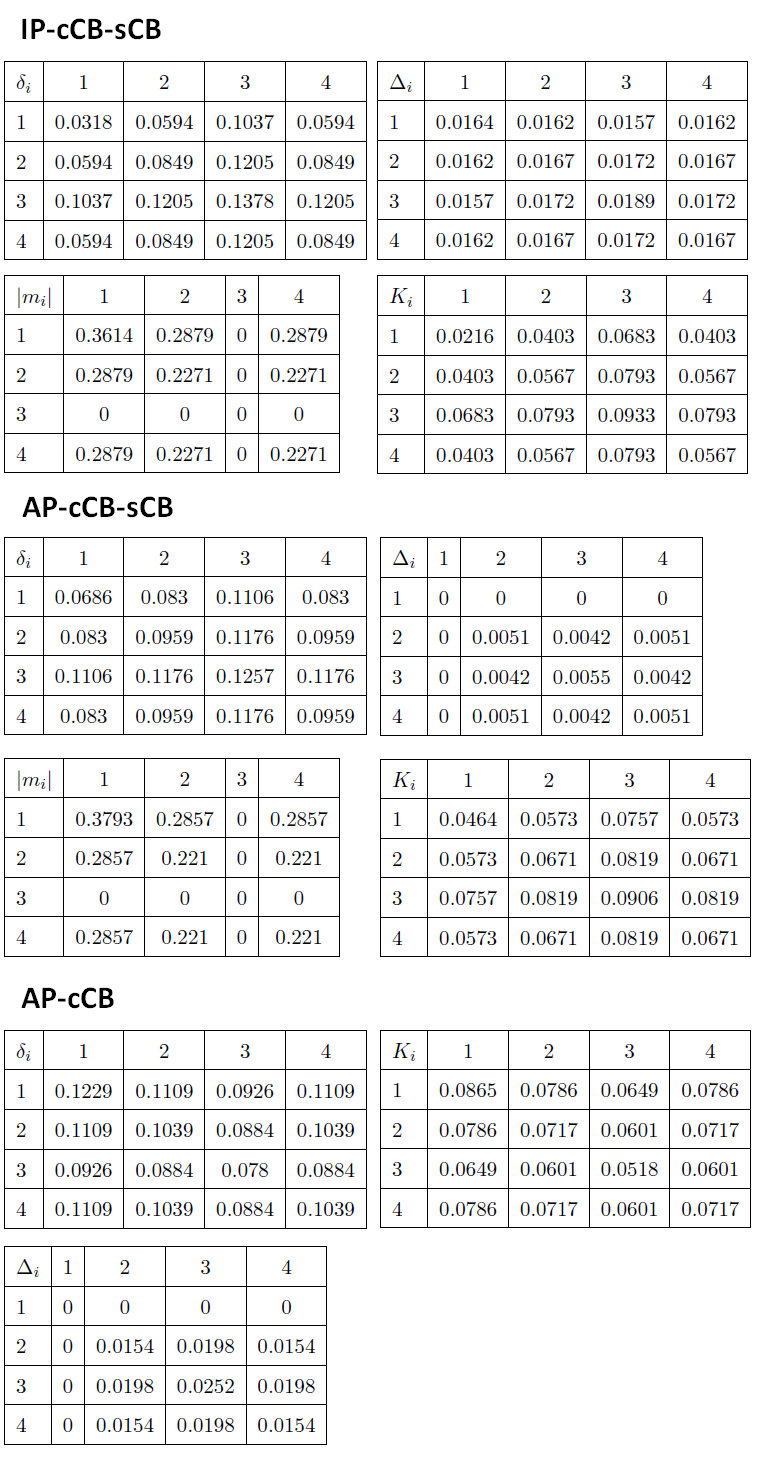}
\caption{Values of several order parameters for checkerboard patterns. The hole concentration is 0.1 for AP-cCB-sCB and AP-cCB but 0.09 for IP-cCB-sCB.}
\label{TabS2}
\end{table}

\begin{figure}[ht]
\centering
\includegraphics[width=6in]{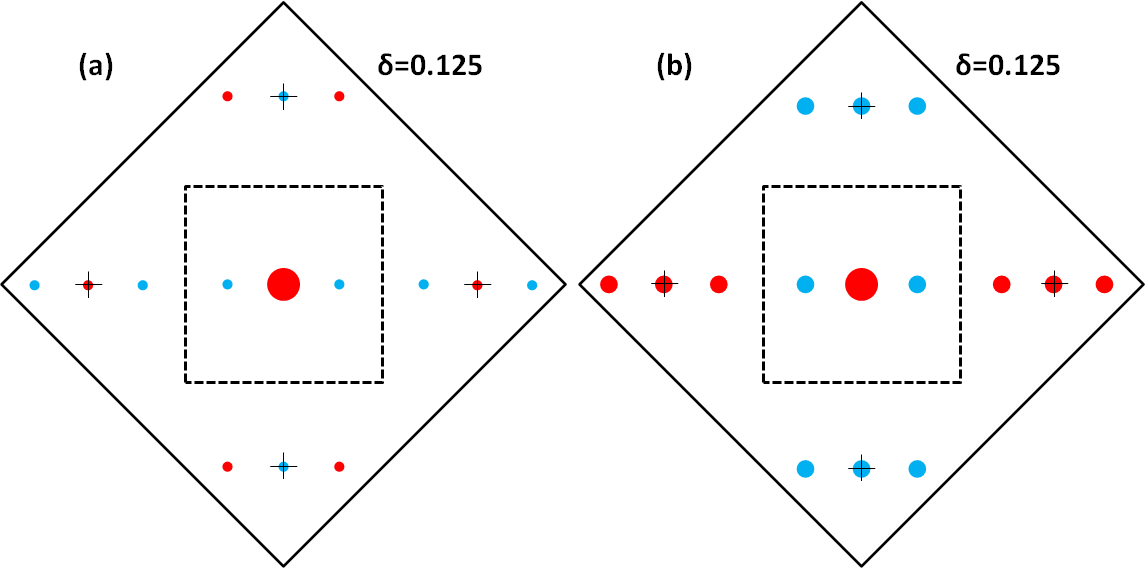}
\caption{ Schematic illustration of the Fourier transform of the bond orders for (a)IP-CDW-SDW stripe and for (b)AP-CDW-SDW. The dot size scales with the magnitude and red (blue) for positive(negative) values.  "+" signs are at the four reciprocal lattice vectors $(\pm2\pi/a_0,0)$ and $(0,\pm2\pi/a_0)$ and their nearby medium size dots are shifted from them by $(\pm\pi/2a_0,0)$.  The center large dot is $Q=(0,0)$ and has two red small dots nearby at $(\pm\pi/2a_0,0)$. The inner dotted square is the boundary of first Brillouin zone. The doping for both stripes is 1/8.}
\label{FigS2}
\end{figure}

\begin{figure}[ht]
\centering
\includegraphics[width=6in]{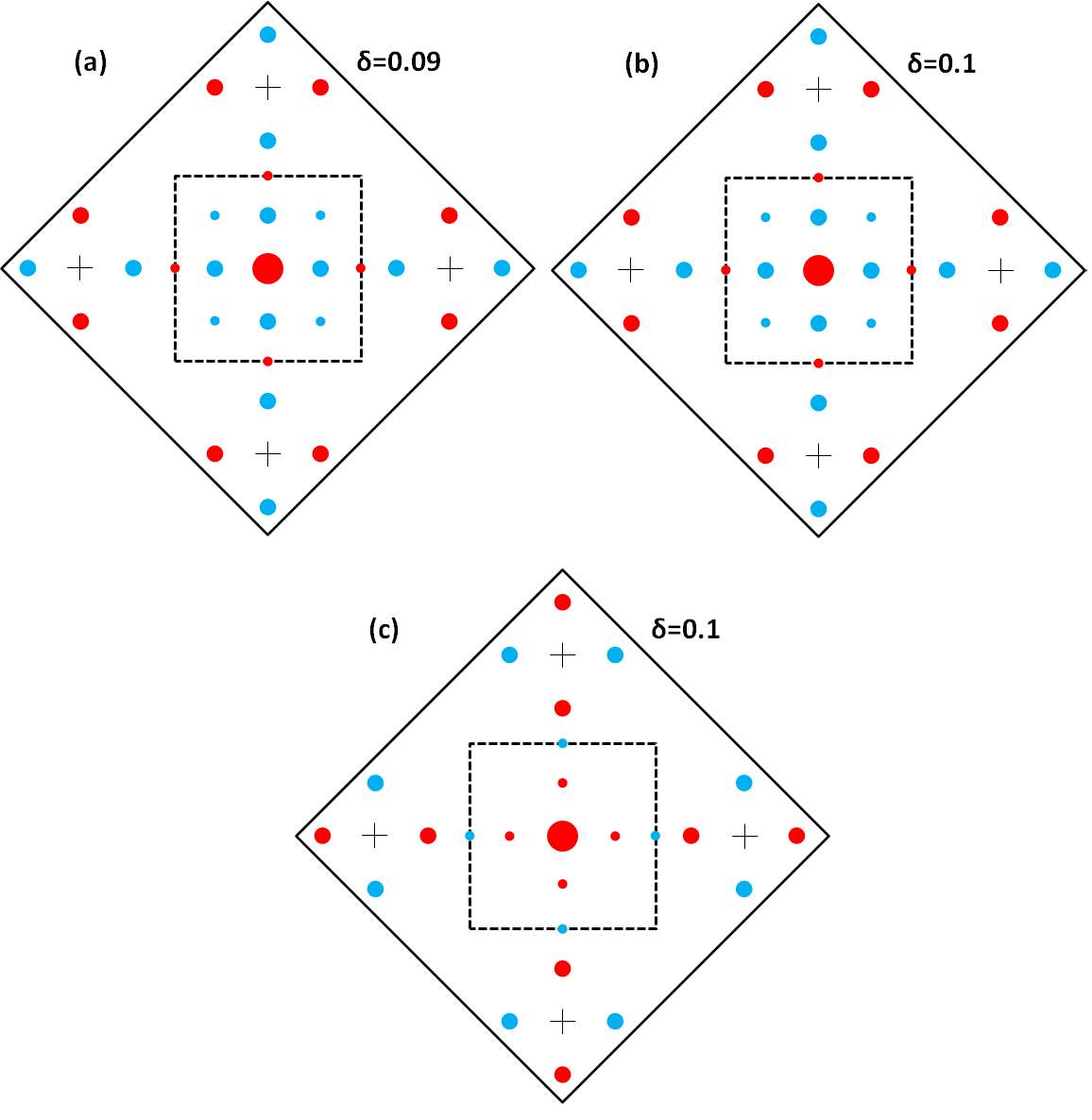}
\caption{Schematic illustration of the Fourier transform  of the bond orders for CB patterns (a)IP-cCB-sCB, (b)AP-cCB-sCB, and (c)AP-cCB. The hole density is (a)$\delta=0.09$, (b)$\delta=0.1$, and (c)$\delta=0.1$. All the dots are shifted from $Q=(0,0)$ and the four reciprocal lattice vectors (denoted by "+" sign) by $(\pm2\pi/a_0,0)$ or $(0,\pm2\pi/a_0)$. The notations are the same as those in Figure 8.}
\label{FigS3}
\end{figure}

\end{document}